\documentclass[prb,epsfigure,twocolumn,showpacs,preprintnumbers,amsmath,amssymb]{revtex4}
\usepackage{graphicx}

\begin{document}

\title{A frustrated nanomechanical device powered by the lateral Casimir force}

\author{MirFaez Miri }
\email{miri@iasbs.ac.ir} \affiliation{Institute for Advanced Studies
in Basic Sciences (IASBS), P. O. Box 45195-1159, Zanjan 45195, Iran}

\author{Ramin Golestanian}
\email{r.golestanian@sheffield.ac.uk} \affiliation{Department of
Physics and Astronomy, University of Sheffield, Sheffield S3 7RH,
UK}

\date{\today}
\begin{abstract}
The coupling between corrugated surfaces due to the lateral Casimir
force is employed to propose a nanoscale mechanical device composed
of two racks and a pinion. The noncontact nature of the interaction
allows for the system to be made frustrated by choosing the two
racks to move in the same direction and forcing the pinion to choose
between two opposite directions. This leads to a rich and sensitive
phase behavior, which makes the device potentially useful as a
mechanical {\em sensor} or {\em amplifier}. The device could also be
used to make a mechanical {\em clock} signal of tunable frequency.
\end{abstract}

\pacs{07.10.Cm,42.50.Lc,46.55.+d,85.85.+j}

\maketitle

With the ongoing miniaturization of the mechanical systems, managing
the tribological interactions at small length scales is proving to
be one of the most significant challenges ahead of us.\cite{Carpick}
A specific area of concern is the durability of mechanical parts
that have fine geometrical features and come in contact with one
another, as they can wear out very quickly.\cite{Wear} Another
important issue is to avoid the stiction of the mechanical
components in small machines.\cite{Buks,recent-stra1,recent-stra2}
These problems suggest that we should try and develop design
strategies for small scale mechanical systems that do not rely
heavily on physical contact between machine parts.

In recent years, the Casimir force\cite{casimir1,casimir2}---that
originates from the quantum fluctuations of the electromagnetic
vacuum---has emerged as a surprise candidate for noncontact
actuation of mechanical devices.\cite{Capasso1,Capasso2} Although
the standard normal Casimir force in the original parallel-plate
geometry might offer limited applicability because it is susceptible
to instabilities,\cite{Buks} it is possible to take advantage of
geometrical features such as corrugations on the surfaces and
produce a lateral component to the Casimir force\cite{golestan97}
that can be used in stable mechanical force
transduction.\cite{Mohideen}

The lateral Casimir force between corrugated surfaces has been
recently used as a basis for designing noncontact mechanical
devices. It has been shown that the lateral Casimir grip can hold up
relatively high velocities in a rack-and-pinion device\cite{amg-prl}
and that such a device can have a rich dynamical phase behavior.
This includes, for example, the possibility of spontaneous symmetry
breaking in the form of the rectification of lateral
vibrations.\cite{amg-pre} It has also been shown that normal
undulations can be rectified into net lateral motion in such devices
using a ratchet-like mechanism.\cite{emig}

\begin{figure}[hbp]
\includegraphics[width=0.85\columnwidth]{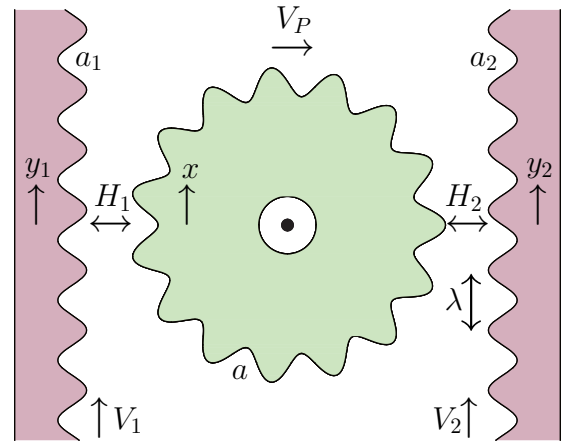}
\caption{(Color online). The schematics of the rack-pinion-rack
device. The choice of parallel rack velocities $V_1$ and $V_2$
(rather than opposite) frustrates the system, which is only possible
because of the noncontact design. The pinion velocity $V_P$ is taken
as positive if it is in the direction shown, as a convention.}
\label{fig:schem}
\end{figure}

Here, we propose a device made with two racks and a pinion that is
coupled to them via the lateral Casimir force, as shown
schematically in Fig. \ref{fig:schem}. The racks are set to move in
the same direction, which renders the dynamics of the pinion
frustrated due to the competing nonlinear couplings. The combination
of frustration and nonlinearity could readily drive the system
towards erratic behavior and chaos, which would be an undesirable
characteristic for a mechanical device. Therefore, we choose to
consider the case of a heavily damped system, so that inertia can be
neglected, which helps avoid the possibility of chaotic behavior. We
probe the motion of the pinion in the different parts of the
parameter space corresponding to the two rack velocities and the two
amplitudes of the lateral Casimir force coupling that depend on the
geometrical characteristics of the setup (see Fig. \ref{fig:schem}).
We find that the system could have five distinct behaviors: (i) the
pinion could be locked with either rack-1 or rack-2, (ii) the pinion
could move along with either rack-1 or rack-2 but with a lower
average velocity, and (iii) the pinion could oscillate back and
forth without choosing to go with either of the racks. The
oscillatory regime could be used to generate a clock signal of
tunable frequency.

\begin{figure}[htp]
\includegraphics[width=0.99\columnwidth]{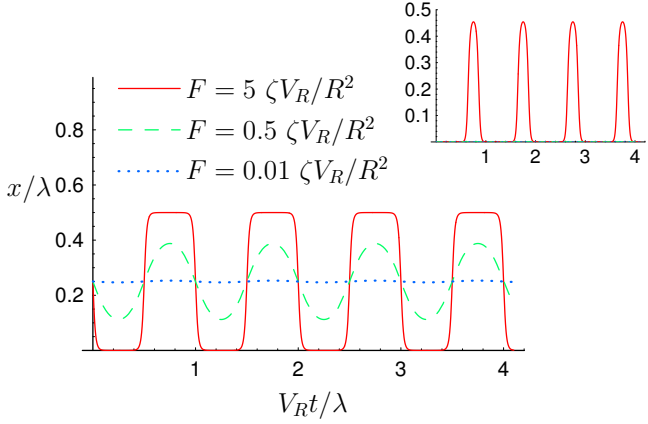}
\caption{(Color online). Displacement of the pinion cogs as a
function of time [from Eq. (\ref{x-t})] for the fully symmetric
device, and $x_0=0.25 \lambda$. For high values of the grip, the
motion resembles a square-wave pattern. Inset: the same (three)
plots for $x_0=0.0001 \lambda$.} \label{fig:clock}
\end{figure}

The corrugated surface of the pinion experiences a lateral Casimir
force of the form\cite{golestan97} $F_{\rm lateral}=F \sin\left[2
\pi (x-y)/\lambda\right]$ from each of the racks, where $x$ and $y$
represent the lateral positioning of the surfaces and $\lambda$ is
the wavelength of the corrugations (see Fig. \ref{fig:schem}), which
must be the same on all three surfaces so that coherent coupling is
possible. The amplitude of the lateral Casimir force, or the
``Casimir grip,'' $F$ depends on the geometric characteristics of
the device and in particular the gap size $H$ and the amplitudes of
corrugations.\cite{Mohideen,Emig-etal-2001,Reynaud,amg-prl,amg-pre}
These forces add up to exert a net torque of
\begin{math}
-R F_1 \sin\left[2 \pi (x-y_1)/\lambda\right]-R F_2 \sin\left[2 \pi
(x+y_2)/\lambda\right]
\end{math}
on the pinion, which should be balanced against the friction torque
to obtain the equation of motion for $x=R \theta$, where $\theta$ is
the angle of rotation and $R$ is the radius of the pinion. Putting
$y_1=V_1 t$ and $y_2=V_2 t$, the equation of motion reads
\begin{equation}
\frac{\zeta}{R^2}\frac{d x}{d t}=-F_1 \sin\left[\frac{2 \pi (x-V_1
t)}{\lambda}\right]-F_2 \sin\left[\frac{2 \pi (x+V_2
t)}{\lambda}\right], \label{master2}
\end{equation}
where $\zeta$ is the rotational friction coefficient.

\begin{figure}[t]
\includegraphics[width=0.99\columnwidth]{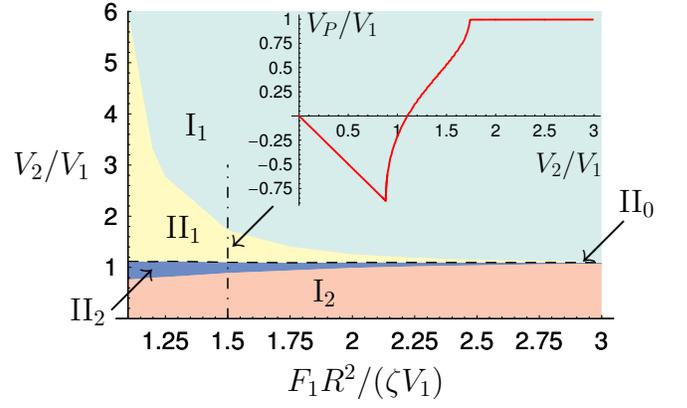}
\caption{(Color online). Phase diagram for the motion of the pinion,
corresponding to $F_2=F_1+0.05 \;\zeta V_1/R^2$. There are five
different behaviors: I$_1$ locked with rack-1, II$_1$ moving with
rack-1 with skipping, II$_0$ neutral oscillatory motion (dashed
line), II$_2$ moving with rack-2 with skipping, and I$_2$ locked
with rack-2. Inset: pinion velocity versus rack velocity for
$F_1=1.50 \;\zeta V_1/R^2$ and $F_2=1.55 \;\zeta V_1/R^2$
(corresponding to a vertical cut of the phase diagram at the
dash-dotted line) showing the five different regimes.}
\label{fig:phase}
\end{figure}

Equation (\ref{master2}) is symmetric under the combined
transformation of $V_1 \leftrightarrow V_2$, $F_1 \leftrightarrow
F_2$, and $x \leftrightarrow -x$. In the fully symmetric case of
$V_R \equiv V_1=V_2$ and $F \equiv F_1=F_2$, the pinion cannot
choose a sense of rotation over the other and performs an
oscillatory motion. In this case, Eq. (\ref{master2}) can be solved
analytically to yield
\begin{equation}
x=\frac{\lambda}{\pi} \tan^{-1} \left\{\tan\left(\frac{\pi
x_0}{\lambda}\right) \exp\left[-\frac{2 R^2 F}{\zeta V_R}
\sin\left(\frac{2 \pi V_R
t}{\lambda}\right)\right]\right\},\label{x-t}
\end{equation}
where $x_0$ is the initial position of the pinion. Note that the
sign of $x_0$ determines the sign of the whole solution $x(t)$.
Equation (\ref{x-t}) is plotted in Fig. \ref{fig:clock} for
different values of the Casimir grip $F$ and the initial pinion
position. It shows that the device can generate interesting periodic
patterns such as a nearly perfect square wave or a train of spikes,
with a frequency $f=V_R/\lambda$ that can be easily controlled by
the rack velocity. The solution [Eq. (\ref{x-t})] involves an
exponential amplification factor that is controlled by the Casimir
grip, which could help detect incredibly small displacements. This
can be seen in the example plotted in the inset of Fig.
\ref{fig:clock}, which shows a $10^4$-fold amplification for
$F=5~\zeta V_R/R^2$.

The general form of Eq. (\ref{master2}) allows for richer behavior
as the pinion can choose to adopt one net sense of rotation over the
other. The pinion can be locked to a rack moving at a given
velocity, if the friction force for such velocities can be balanced
by the available Casimir grip. This means that the Casimir grips
$F_1$ and $F_2$ introduce two {\em skipping} velocities $V_{S1} \sim
F_1 R^2/\zeta$ and $V_{S2} \sim F_2 R^2/\zeta$, which determine that
for $V_1 < V_{S1}$ and $V_2 < V_{S2}$ locking of the pinion to the
corresponding rack is possible. For larger velocities, the pinion
and the corresponding rack will skip cogs.

We have studied the behavior of the device numerically in various
parts of the large parameter space of the system. Figure
\ref{fig:phase} shows a representative phase diagram for a specific
section of the space of parameters. It can be seen that the system
exhibits five different behaviors ranging from the pinion being
locked to rack-1 (I$_1$) or rack-2 (I$_2$) to motion with skipping
along rack-1 (II$_1$) or rack-2 (II$_2$). The phase boundary between
the two II-phases (shown in Fig. \ref{fig:phase} as a dashed line
and denoted as II$_0$) corresponds to the vanishing of the net
pinion velocity and thus an oscillatory behavior similar to the
symmetric device as described by the solution in Eq. (\ref{x-t}) and
Fig. \ref{fig:clock}. Note that the phase boundary between I$_1$ and
II$_1$ in Fig. \ref{fig:phase} is about to asymptote to a vertical
line at lower values of $F_1$ below which the phase I$_1$ cannot
exist (corresponding to $V_{S1}$ discussed above). The inset of Fig.
\ref{fig:phase} shows the average pinion velocity for a typical
section of the phase diagram, going from $V_P=-V_2$ to $V_P=V_1$
passing through two sharp transition points and zero.

The assumption of a heavily damped system is not unrealistic in view
of the technical difficulties of mounting the pinion on an axle or
pivot. We can estimate the value of the rotational friction
coefficient $\zeta$ assuming that the main source of friction in the
system comes from the lubrication at the axle. For an axle of radius
$r$ which is lubricated with a fluid layer of thickness $h$ and
viscosity $\eta$, we find $\zeta \simeq 2 \pi \eta L r^3/h$ where
$L$ is the thickness (or height) of the pinion. We can also estimate
the moment of inertia of the pinion assuming it is a cylinder of
mass $M$ and density $\rho$, as $I=\frac{1}{2} M R^2=\frac{\pi}{2}
\rho L R^4$. The characteristic time scale that probes the relative
importance of inertia and friction is defined as $\tau=I/\zeta=\rho
h R^4/(4 \eta r^3)$. Using $\rho=1.17$ gr/cm$^3$ (for silicone),
$\eta=10^{-3}$ Pa.s (for a lubricant as thick as water), and the
geometrical parameters as $R=1 \;\mu{\rm m}$, $L=10 \;\mu{\rm m}$,
$r=500 \;{\rm nm}$, and $h=100 \;{\rm nm}$, we find $\tau=2.3 \times
10^{-7}$ s. For time scales bigger than that, the assumption of a
heavily damped system is reasonable.

We can also estimate the skipping velocities using typical values
for the Casimir grip, whose value is very sensitive to the gap
size.\cite{Mohideen,Emig-etal-2001,Reynaud,amg-prl,amg-pre} For
typical (and experimentally realized\cite{mohideen2}) values of
$a=50 \;{\rm nm}$ (assumed for all surfaces) and $\lambda=500 \;{\rm
nm}$, we find $F=0.3 \;{\rm pN}$ for $H=200 \;{\rm nm}$ that yields
a skipping velocity of $V_{S} \sim F R^2/\zeta=3.8 \;\mu{\rm m/s}$.
However, reducing the gap size (by only a factor of {\bf two}) to
$H=100 \;{\rm nm}$ yields $F=12 \;{\rm pN}$ and consequently $V_{S}
\sim F R^2/\zeta=150 \;\mu{\rm m/s}$ (that is enhanced by factor of
{\bf forty}). The estimates for the amplitude $F$ have been made
assuming the boundaries are perfect metals. In practical situations
with real metallic boundaries of smaller reflectivity, the Casimir
grip tends to be slightly weaker, but the general behavior of the
system will still be the same.

The clock signal does not necessarily need to come from the fully
symmetric device, and as the phase diagram of Fig. \ref{fig:phase}
shows it can be obtained by tuning the system into the dashed line
(II$_0$) transition boundary for any geometrical design. To get a
strong coupling signal like the square wave or the train of spikes
shown in Fig. \ref{fig:clock}, we need to have a rack velocity of
say $V_R=F R^2/(5 \zeta)$, which is equal to $30 \;\mu{\rm m/s}$ for
$H=100 \;{\rm nm}$. This yields $f=V_R/\lambda=60$ Hz using the
above value for the corrugation wavelength. Higher rack velocities
can lead to higher frequencies, although the shape of the signal
will change to a simple sinusoidal one as the coupling gets weaker
progressively with increasing rack velocity.

In conclusion, we have proposed a design for a mechanical device
made of a nanoscale pinion sandwiched without contact between two
racks that exert opposing forces, by employing the quantum
fluctuations of the electromagnetic field. Because of the
frustration that is built into the design, the system can react
dramatically to minute changes in the geometrical features in the
system and can thus act as a good sensor. The noncontact nature of
this device, and other variants based on similar principles, could
help us in our quest towards wearproof nanoscale mechanical
engineering.

%\acknowledgments

It is a pleasure to acknowledge fruitful discussions with R.A.L.
Jones. This work was supported by EPSRC under Grant EP/E024076/1.


\begin{thebibliography}{99}

\bibitem{Carpick}
R.W. Carpick, Science {\bf 313}, 184 (2006).

\bibitem{Wear}
M.P. de Boer, T.M. Mayer, MRS Bull. {\bf 26}, 302 (2001).

\bibitem{Buks}
E. Buks and M.L. Roukes, Phys. Rev. B {\bf 63}, 033402 (2001).
%;Nature {\bf 419}, 119 (2002).

\bibitem{recent-stra1}
A. Socoliuc, E. Gnecco, S. Maier, O. Pfeiffer, A. Baratoff, R.
Bennewitz, E. Meyer, Science {\bf 313}, 207 (2006).

\bibitem{recent-stra2}
J.Y. Park, D.F. Ogletree, P.A. Thiel, and M. Salmeron, Science {\bf
313}, 186 (2006).

\bibitem{casimir1}
H.B.G. Casimir, Proc. K. Ned. Akad. Wet. {\bf 51}, 793 (1948).

\bibitem{casimir2}
M. Bordag, U. Mohideen, and V. M. Mostepanenko, Phys. Rep. {\bf
353}, 1 (2001).

\bibitem{Capasso1}
H.B. Chan, V.A. Aksyuk, R.N. Kleiman, D.J. Bishop and F. Capasso,
Science {\bf 291}, 1941 (2001).

\bibitem{Capasso2}
H.B. Chan, V.A. Aksyuk, R.N. Kleiman, D.J. Bishop and F. Capasso,
Phys. Rev. Lett. {\bf 87}, 211801 (2001).

\bibitem{golestan97}
R. Golestanian and M. Kardar, Phys. Rev. Lett. {\bf 78}, 3421
(1997); Phys. Rev. A {\bf 58}, 1713 (1998).

\bibitem{Mohideen}
F. Chen, U. Mohideen, G.L. Klimchitskaya, and V.M. Mostepanenko,
Phys. Rev. Lett. {\bf 88}, 101801 (2002); Phys. Rev. A {\bf 66},
032113 (2002).

\bibitem{amg-prl}
A. Ashourvan, M.F. Miri, and R. Golestanian, Phys. Rev. Lett. {\bf
98}, 140801 (2007).

\bibitem{amg-pre}
A. Ashourvan, M.F. Miri, and R. Golestanian, Phys. Rev. E {\bf 75},
040103 (R) (2007).

\bibitem{emig}
T. Emig, Phys. Rev. Lett. {\bf 98}, 160801 (2007).

\bibitem{Emig-etal-2001}
T. Emig, A. Hanke, R. Golestanian, and M. Kardar, Phys. Rev. Lett.
{\bf 87}, 260402 (2001); Phys. Rev. A {\bf 67}, 022114 (2003).

\bibitem{Reynaud}
R.B. Rodrigues, P.A. Maia Neto, A. Lambrecht, and S. Reynaud, Phys.
Rev. Lett. {\bf 96}, 100402 (2006).

\bibitem{mohideen2}
U. Mohideen, private communication.

\end{thebibliography}
\end{document}